\documentclass[12pt]{article}
\usepackage{graphicx}
\usepackage{cite}
\usepackage{amsmath} 
\usepackage{hyperref}
\input{epsf} 
\def\ltap{\raisebox{-.6ex}{\rlap{$\,\sim\,$}} \raisebox{.4ex}{$\,<\,$}} 
\def\gtap{\raisebox{-.6ex}{\rlap{$\,\sim\,$}} \raisebox{.4ex}{$\,>\,$}}

\newcommand\as{\alpha_{\mathrm{S}}} 
 
\def\beq{\begin{equation}} 
\def\eeq{\end{equation}} 
\def\beeq{\begin{eqnarray}} 
\def\eeeq{\end{eqnarray}}

\def\to{\rightarrow}

\def\WH{{\it WH\,}}
\def\WpH{{\it W$\;\!^+\!$H\,}}
\def\WmH{{\it W$\;\!^-\!$H\,}}
\def\ZH{{\it ZH\,}}
\def\VH{{\it VH\,}}

\begin{document} 

\begin{titlepage}
\begin{flushright}
TIF-UNIMI-2017-5\\
\end{flushright}
\renewcommand{\thefootnote}{\fnsymbol{footnote}}
\par \vspace{10mm}

\begin{center}
{\Large \bf Associated production of a Higgs boson decaying 
}
\\[0.5cm]
{\Large \bf into 
bottom quarks at the LHC in full NNLO QCD}
\end{center}
\par \vspace{2mm}
\begin{center}
{\bf Giancarlo Ferrera}$^{(a)}$, {\bf G\'abor Somogyi}$^{(b)}$
~and~~{\bf Francesco Tramontano}$^{(c)}$\\

\vspace{5mm}

$^{(a)}$ TIF Lab, Dipartimento di Fisica, Universit\`a di Milano and\\
INFN, Sezione di Milano, I-20133 Milan, Italy\\
$^{(b)}$ MTA-DE Particle Physics Research Group, \\
H-4010 Debrecen, PO Box 105, Hungary\\
$^{(c)}$ Dipartimento di Fisica, Universit\`a di Napoli and\\
INFN, Sezione di Napoli, I-80126 Naples, Italy

\vspace{5mm}

\end{center}

\par \vspace{2mm}
\begin{center} {\large \bf Abstract} \end{center}
\begin{quote}
\pretolerance 10000

We consider the production of a Standard Model Higgs boson decaying to bottom quarks
in association with a vector boson $W^\pm/Z$ in hadron collisions. We present
a fully exclusive calculation of QCD radiative corrections 
both for the production cross section and for the Higgs boson decay rate 
up to next-to-next-to-leading order (NNLO) accuracy.
Our calculation also includes the leptonic decay of the vector boson 
with finite-width effects and spin correlations. 
We consider typical kinematical cuts applied in the experimental analyses at the Large Hadron Collider (LHC)
and we find that the {\itshape full} NNLO QCD corrections significantly decrease the accepted cross section and 
have a substantial impact on the shape of distributions.
We point out that these additional effects are essential to obtain precise theoretical predictions
to be compared with the LHC data.

\end{quote}

\vspace*{\fill}
\begin{flushleft}
May 2017

\end{flushleft}
\end{titlepage}

\setcounter{footnote}{1}
\renewcommand{\thefootnote}{\fnsymbol{footnote}}

The discovery of the long sought Higgs boson 
($H$)~\cite{Higgs:1964ia,Englert:1964et} by the ATLAS and CMS Collaborations 
at the  Large Hadron Collider (LHC)~\cite{Aad:2012tfa,Chatrchyan:2012ufa} paved the way for the
experimental investigation of the electroweak (EW) symmetry breaking mechanism and, in particular,
for the measurements of the Higgs boson couplings to the Standard Model (SM) particles. 
In this respect the increasing amount of precise experimental data collected at the LHC demands 
a corresponding  improvement of theoretical predictions.

One of the main production mechanisms of the SM Higgs boson at hadron colliders is the
associated production with a vector boson ($V=W^\pm,Z$).
This process offers the unique opportunity to study both the Higgs boson coupling to massive gauge bosons and
to bottom ($b$) quarks via the decay $H\to b\bar{b}$. 

A direct search for the SM Higgs boson through associated \VH\ production and  $H\to b\bar{b}$ decay
has been carried out at the LHC at a centre--of--mass energy 
of $\sqrt{s}=7/8$~TeV~\cite{Aad:2014xzb,Chatrchyan:2013zna} and at $\sqrt{s}=13$~TeV~\cite{Aaboud:2017xsd,Sirunyan:2017elk}.
The ATLAS and CMS Collaborations observed an excess of events above the expected background with a 
measured signal strength relative to the SM expectation of $0.90 \pm 0.18 \mathrm{(stat.)} ^{+0.21}_{-0.19} \mathrm{(syst.)}$~\cite{Aaboud:2017xsd}
and $\mu=1.06^{+0.31}_{-0.29}$\cite{Sirunyan:2017elk} respectively.

High precision theoretical predictions require detailed computations of 
radiative corrections for cross sections and corresponding
distributions. 
The total cross section for associated \VH\ production is known 
at next-to-next-to-leading
order (NNLO) in 
QCD~\cite{Hamberg:1990np,Brein:2003wg,Brein:2011vx,Brein:2012ne} 
and next-to-leading order (NLO) in the electroweak
theory~\cite{Ciccolini:2003jy,Denner:2011id}. 
Fully differential calculations have been performed
in NNLO QCD for the \VH\ production cross section, together with the NLO QCD corrections for the 
Higgs boson decay rate into bottom 
quarks~\cite{Ferrera:2011bk,Ferrera:2013yga,Ferrera:2014lca,Campbell:2016jau}. 
The fully differential $H\to b{\bar b}$ decay rate has been computed up to 
NNLO in QCD~\cite{Anastasiou:2011qx,DelDuca:2015zqa} 
while the inclusive rate is known up to 
${\mathcal O}(\alpha_S^4)$~\cite{Baikov:2005rw} and up to NLO in the electroweak theory~\cite{Kniehl:1991ze,Dabelstein:1991ky}.
Resummation and higher order (beyond NNLO) QCD effects have been investigated in 
Refs.\cite{Dawson:2012gs,Altenkamp:2012sx,Li:2014ria,Shao:2013uba,Harlander:2014wda,Kumar:2014uwa,Hasselhuhn:2016rqt}
while  the combination of fixed-order QCD calculations with parton shower Monte Carlo algorithms 
has been considered in Refs.~\cite{Hamilton:2012rf,Luisoni:2013cuh,Astill:2016hpa}.
NLO EW effects at fully differential level for \VH\ production with the leptonic decay of the vector boson
have been considered in Refs.~\cite{Denner:2011id,Denner:2014cla}.

In this letter we present the fully differential calculation of the 
NNLO QCD corrections for the production cross section 
{\itshape and} for the Higgs boson decay rate to bottom quarks,
exploiting the very good accuracy of the narrow width approximation for the Higgs boson ($\Gamma_H\ll m_H$).
In Refs.~\cite{Banfi:2012jh,Ferrera:2013yga} it was shown that,
when the set of kinematical cuts
applied in the LHC analyses are considered,
the effect of QCD corrections to the Higgs
boson decay process can be large.
Motivated by these findings, we extend existing calculations on higher order QCD predictions 
by considering the {\itshape complete} second order terms.  Together with the
NNLO corrections for the production cross section, we include
the NNLO corrections to the  $H\to b\bar{b}$ decay
rate {\itshape and} the combination of the NLO 
terms for the production and decay stages. 

We implemented our computation in the parton level Monte Carlo numerical 
program {\ttfamily HVNNLO} which allows the user to apply arbitrary kinematical cuts on final-state leptons,  
$b$ jets  and associated QCD radiation, and to compute the corresponding distributions in the form of histograms.

The main result of our study is that 
for a typical set of kinematical cuts applied in the LHC analyses
 we observe a substantial decrease of the {\itshape complete} NNLO QCD 
prediction with respect to lower order calculations. 
Therefore the inclusion of the QCD effects we have calculated 
could be relevant to improve the agreement between the SM predictions 
and the current LHC data.
In this letter we discuss the main ingredients of our computation,
a more comprehensive phenomenological analysis will appear elsewhere.

We consider the inclusive hard scattering reaction
$\,h_1+h_2 \rightarrow VH+X \rightarrow l_1l_2 b \bar{b} +X\,$,
where the collision of the hadrons $h_1$ and $h_2$ produces the $\VH$ system ($V=W^\pm,Z$) which
subsequently decays into the lepton pair $l_1l_2$ ($l_1l_2\!\equiv\! l\nu_l$ in the case of $W^\pm$ decay) 
and the bottom quark-antiquark pair $b \bar{b}$, while $X$ denotes the accompanying
QCD radiation. We consider a high value of the $\VH$ invariant mass ($M_{\VH}$), which sets the hard-scattering scale of the process, 
and we treat the colliding
hadrons, the leptons and the $b$ quarks in the massless approximation.

By using the narrow width approximation for the Higgs boson, 
the perturbative QCD expansion of the fully differential cross section 
can be written in the following factorized form\,\footnote{In order to simplify the notation 
the leptonic decay $V\to l_1l_2$  of the $V$ boson (including spin correlations) has been understood, since it has no effect  
from the point of view of QCD corrections.}:
\begin{equation}
\label{expand}
d\sigma_{h_1h_2 \to V\!H \to V b{\bar b}}=
d\sigma_{h_1h_2 \to V\!H} \times \frac{d\Gamma_{H \rightarrow b\bar{b}}}{\Gamma_{H}}=
\left[\sum_{k=0}^{\infty} d\sigma^{(k)}_{h_1h_2 \rightarrow \VH} \right] \times
\left[\frac{\sum_{k=0}^{\infty} d\Gamma^{(k)}_{H \rightarrow b\bar{b}}}
{\sum_{k=0}^{\infty} \Gamma^{(k)}_{H \rightarrow b\bar{b}}} \right] \times
{\mbox{Br}}(H \rightarrow b\bar{b})\,,
\end{equation}
where $\Gamma_{H \rightarrow b\bar{b}}$ and $\Gamma_H$ are the Higgs boson partial decay width to bottom quarks and the total decay width
respectively, and the expansion in powers of $\alpha_S$ is given by the exponent $k$.
Eq.\,(\ref{expand}) is arranged in a form such that we can exploit the precise prediction of the Higgs boson branching ratio into
$b$ quarks  $\mbox{Br}(H \rightarrow b\bar{b})=\Gamma_{H \rightarrow b\bar{b}}/\Gamma_{H}$  
(see for instance Ref.~\cite{Dittmaier:2011ti}), by which 
we normalize the contributions to the differential decay rate of the Higgs boson\,\footnote{Indeed, by considering observables that
are inclusive over the Higgs boson decay products, we obtain the production cross section times the branching ratio 
$d\sigma_{h_1h_2 \to V\!H} \times {\mbox{Br}}(H \rightarrow b\bar{b})$.}.

By expanding Eq.\,(\ref{expand}) up to the second order in $\alpha_S$ we have:
\begin{eqnarray}
\label{expand2}
d
\sigma^{\rm NNLO}_{h_1h_2 \to V\!H \to V b{\bar b}}&=& 
\left[ d\sigma^{(0)}_{h_1h_2 \rightarrow \VH} \times
\frac{d\Gamma^{(0)}_{H \rightarrow b\bar{b}}
    + d\Gamma^{(1)}_{H \rightarrow b\bar{b}}
    + d\Gamma^{(2)}_{H \rightarrow b\bar{b}}}
     {\Gamma^{(0)}_{H \rightarrow b\bar{b}}
    + \Gamma^{(1)}_{H \rightarrow b\bar{b}}
    + \Gamma^{(2)}_{H \rightarrow b\bar{b}}}\right.
\\\nonumber
&+& 
\left.
d\sigma^{(1)}_{h_1h_2 \rightarrow \VH} \times
\frac{d\Gamma^{(0)}_{H \rightarrow b\bar{b}}
    + d\Gamma^{(1)}_{H \rightarrow b\bar{b}}}
     {\Gamma^{(0)}_{H \rightarrow b\bar{b}}
    + \Gamma^{(1)}_{H \rightarrow b\bar{b}}}\right.
\\\nonumber
&+&
\left.
d\sigma^{(2)}_{h_1h_2 \rightarrow \VH} \times
\frac{d\Gamma^{(0)}_{H \rightarrow b\bar{b}}}
     {\Gamma^{(0)}_{H \rightarrow b\bar{b}}}
\right] \times {\mbox{Br}}(H \rightarrow b\bar{b})\,.
\end{eqnarray}
Eq.\,(\ref{expand2}) contains the {\itshape complete} NNLO contributions,
which include the $\alpha_S^2$ corrections: {\itshape (i)} to the decay rate (included in the first term on the right hand side),
{\itshape (ii)} from the combination of the NLO contributions for production and decay 
(included in the second term on the right hand side), and
{\itshape (iii)} to the production cross section (the third term on the right hand side).
The novel result of this letter compared with previous approximations
concerns the full computation of the terms  {\itshape (i)} and {\itshape (ii)}. 

In order to compare with the {\itshape partial} NNLO calculations considered so far~\cite{Ferrera:2013yga,Ferrera:2014lca,Campbell:2016jau}, 
we also consider the  truncation of Eq.\,(\ref{expand}) defined as: 
\begin{eqnarray}
\label{expand3}
d
\sigma^{\rm NNLO(prod)+NLO(dec)}_{h_1h_2 \to V\!H \to V b{\bar b}}&=& 
\left[ d\sigma^{(0)}_{h_1h_2 \rightarrow \VH} \times
\frac{d\Gamma^{(0)}_{H \rightarrow b\bar{b}}
    + d\Gamma^{(1)}_{H \rightarrow b\bar{b}}}
     {\Gamma^{(0)}_{H \rightarrow b\bar{b}}
    + \Gamma^{(1)}_{H \rightarrow b\bar{b}}}\right.
\\\nonumber
&+& 
\left.
\left(d\sigma^{(1)}_{h_1h_2 \rightarrow \VH}+d\sigma^{(2)}_{h_1h_2 \rightarrow \VH} \right) \times
\frac{d\Gamma^{(0)}_{H \rightarrow b\bar{b}}}
     {\Gamma^{(0)}_{H \rightarrow b\bar{b}}}
\right] \times {\mbox{Br}}(H \rightarrow b\bar{b})\,,
\end{eqnarray}
which contains the NNLO corrections for the production cross section together with the NLO corrections for the $H\to b{\bar b}$ decay rate.

Each term in the Eqs. (\ref{expand2}) and (\ref{expand3}) includes 
all the relevant contributions from 
{\itshape (double-) real}, {\itshape real-virtual}
and {\itshape (double-) virtual} corrections.
In our implementation we have employed the $q_T$ subtraction method  for the \VH\ production 
cross section~\cite{Ferrera:2011bk,Ferrera:2014lca} and the 
{\itshape CoLoRFulNNLO} method  for the $H\to b{\bar b}$ decay rate~\cite{DelDuca:2015zqa}. 
Details of these formalisms can be found in Refs.\cite{Catani:2007vq,Catani:2013tia} and \cite{Somogyi:2006da,Somogyi:2006db,DelDuca:2016ily} respectively.

While the full NLO and part of the NNLO QCD corrections to \VH\ production are the same as those of the Drell--Yan (DY) 
process~\cite{Catani:2009sm}, with the Higgs boson radiated by the $V$ boson,
additional  contributions appear at NNLO, with the Higgs boson coupled to a heavy-quark loop. 
In the case of \ZH\ production at the LHC, the impact of the gluon-gluon initiated subprocess involving a heavy-quark loop is substantial, due to the large gluon 
luminosity.  We have taken into account these corrections with the full dependence on the top and bottom heavy-quark masses~\cite{Ferrera:2014lca}. 
At NNLO there is yet another set of non DY like contributions
involving quark induced heavy-quark loops both for \ZH\ and \WH\ production.
These corrections have been computed in Ref.~\cite{Brein:2011vx}, relying in some cases on the large-$m_t$ approximation, 
and have been shown to have an impact on the \VH\ cross  section  at the LHC   
at the $1\%$ level (for $m_H\sim 125$\,GeV).  However since the validity of the large-$m_t$ approximation is challenged in the high invariant mass
region probed by the \VH\ kinematics, 
we considered in our computation only the terms which can be presently calculated retaining the full $m_t$ dependence. 
In particular we included  the NNLO terms obtained by radiating the Higgs boson off a top-quark bubble-insertion into an external
gluon line. These terms, called $R_I$ in Ref.~\cite{Brein:2011vx}, contribute both to \ZH\ and \WH\ 
production. On the other hand the $R_{II}$ terms of Ref.~\cite{Brein:2011vx}, which are present only for \ZH\ production,
have been shown to contribute at the sub-per-mille level and have been thus neglected in this paper.  

We are 
interested in the identification of the $b$-quark jets which originate from
the Higgs boson decay.
Besides the $b$-quark pair directly produced in the Higgs boson decay, 
we consistently include the effect of $b$-quark emissions from initial and final state partons\,\footnote{
Therefore, within our NNLO calculation, we have up to four $b$ quarks in the final state.}.
However the standard jet clustering algorithms~\cite{Catani:1993hr} do not
provide an infrared and collinear safe definition of flavoured jets with massless quarks. 
In the present case, at 
NNLO, the splitting of a gluon in a soft or collinear (massless) $b\bar{b}$ pair may affect the 
{\itshape flavour} of a jet. While the collinear unsafety can be removed by defining  as a ``$b$-jet'' a jet containing
a number of $b$ quarks different from the number of $\bar{b}$ quarks, the definition of infrared safe $b$-jets
using standard jet clustering algorithms is less trivial.
In order to deal with an infrared and collinear safe $b$-jet definition, we consider
the so called flavour $k_T$ algorithm~\cite{Banfi:2006hf}. According to this algorithm, the definition of the $k_T$-distance measure 
in the presence of flavoured partons (particles)  is modified in such a way that the flavour of a jet is insensitive to soft 
parton emissions.

We now present numerical results for $pp$ collisions at a center--of--mass energy of $\sqrt{s}=13$~TeV. For
the electroweak couplings, we use the $G_\mu$ scheme and the following input parameters: $G_F = 1.1663787\times 10^{-5}$~GeV$^{-2}$,
$m_Z = 91.1876$~GeV, $m_W = 80.385$~GeV,
$\Gamma_Z=2.4952$~GeV, $\Gamma_W=2.085$~GeV, $m_t=172$~GeV and $m_b=4.18$~GeV\,\footnote{We consider the pole mass
for the top quark ($m_t$) and the $\overline{MS}$ scheme for the bottom quark mass $m_b=\overline{m}_b(\overline{m}_b)$.}.
The mass and the width of the SM Higgs boson are set to $m_H=125$~GeV and $\Gamma_H=4.070$~MeV respectively, while
the $H\to b{\bar b}$ branching ratio  is set to ${\mbox{Br}}(H\to b{\bar b})=0.578$~\cite{Dittmaier:2011ti}.

As for the parton distribution functions (PDFs), we use the NNLO PDF4LHC set~\cite{pdf1} with $\as(m_Z)=0.118$.
We set the renormalization and factorization scales to the dynamical value 
$\mu_R=\mu_F=M_{\VH}$ (i.e.\ the invariant mass of the \VH\ system) and the renormalization scale for the 
$H\to b{\bar b}$ coupling to the value $\mu_r=m_H$.
To assess the impact of scale variation,
we fix $\mu_r=m_H$ varying $\mu_R$ and $\mu_F$ independently in the range 
$M_{VH}/2 \leq \{\mu_R, \mu_F\} \leq 2M_{VH}$, with the constraint $1/2\leq \mu_R/\mu_F\leq 2$. We then fix $\mu_R=\mu_F=M_{VH}$ and vary the decay 
renormalisation scale $\mu_r$ between $m_H/2$ and $2 m_H$. The final uncertainty is obtained by taking the envelope of the two 
(production and decay) scale uncertainties.
Jets are reconstructed with the flavour-$k_T$ algorithm with $R=0.5$~\cite{Banfi:2006hf}. We define a $b$-jet
as a jet which contains a number of $b$ quarks different from the number of anti-$b$ quarks ($N(b)\neq N(\bar{b})$).


\begin{table}[htbp]
\begin{center}
\begin{tabular}{|c|c|c|c|}
\hline
$\sigma$ (fb) 
& NNLO(prod)+NLO(dec) & full NNLO  \phantom{\big\}} \\
\hline
\hline
$pp\to W^+H+X\to l\nu_l b{\bar b}+X$\ & $3.94^{+1\%}_{-1.5\%}$ 
& $3.70^{+1.5\%}_{-1.5\%}$  
\phantom{\Big\}}\\
\hline
$pp\to ZH+X\to \nu\nu b{\bar b}+X$\ & $8.65^{+4.5\%}_{-3.5\%}$  
& $8.24^{+4.5\%}_{-3.5\%}$ 
\phantom{\Big\}} \\ 
\hline
\end{tabular}
\end{center}
\caption{
{\em Cross sections and their scale uncertainties 
for $pp\to VH+X\to l_1l_2b{\bar b}+X$ at LHC with $\sqrt{s}=13$~TeV. The applied kinematical cuts are described in the text.
}}
\label{table1}
\end{table}

We start the presentation of our results by considering \WpH\ production and decay at the LHC at $\sqrt{s}=13$~TeV. 
Our choice of kinematical selection cuts on the final states 
closely follows the fiducial setup considered in the {\itshape CERN Yellow Report}
of the {\itshape LHC Higgs Cross Section Working Group}~\cite{deFlorian:2016spz}.
We require the charged lepton to have transverse momentum $p_T^l > 15$~GeV and pseudorapidity $|\eta_l|< 2.5$ while 
the missing transverse energy of the event is required to be $E_T^{miss}>30$~GeV.
The $W$ boson is required to have a transverse momentum $p_T^W > 150$~GeV.
Finally we require at least two $b$-jets each with $p_T^b > 25$~GeV and $|\eta_b|<2.5$. 
The corresponding cross sections in the fiducial region are reported in the first row of 
Table~\ref{table1},
where we present the  {\itshape full} NNLO prediction (see Eq.\,(\ref{expand2}))
compared with the {\itshape partial} NNLO prediction 
(see Eq.\,(\ref{expand3}))\,\footnote{The results for the case of \WmH\ production and decay 
are qualitative similar, with a numerical reduction of fiducial cross section 
around $40\%$.}. 
We observe that 
the inclusion of the {\itshape full} NNLO corrections reduces the cross section by around $6\%$ with respect to the {\itshape partial} 
NNLO result\,\footnote{In particular we note that roughly $40\%$ of the reduction is due to the combination of the NLO contributions 
for production and decay and $60\%$ is due to the NNLO contributions to the decay rate (see Eq.\.(\ref{expand2}) and subsequent comments).}.

\begin{figure}[th]
\begin{center}
\begin{tabular}{cc}
\includegraphics[width=.45\textwidth]{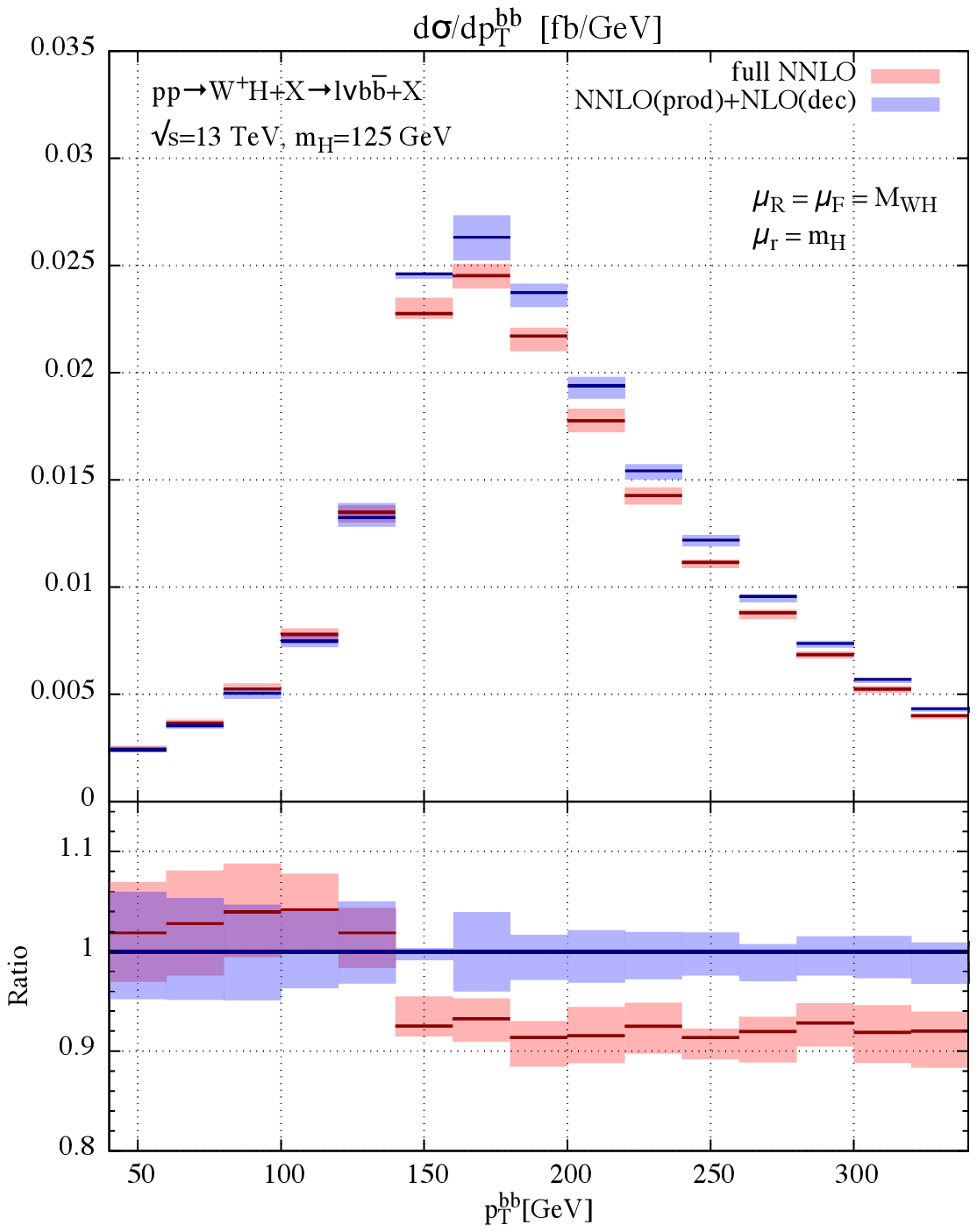}
\hspace*{.75cm}
\includegraphics[width=.45\textwidth]{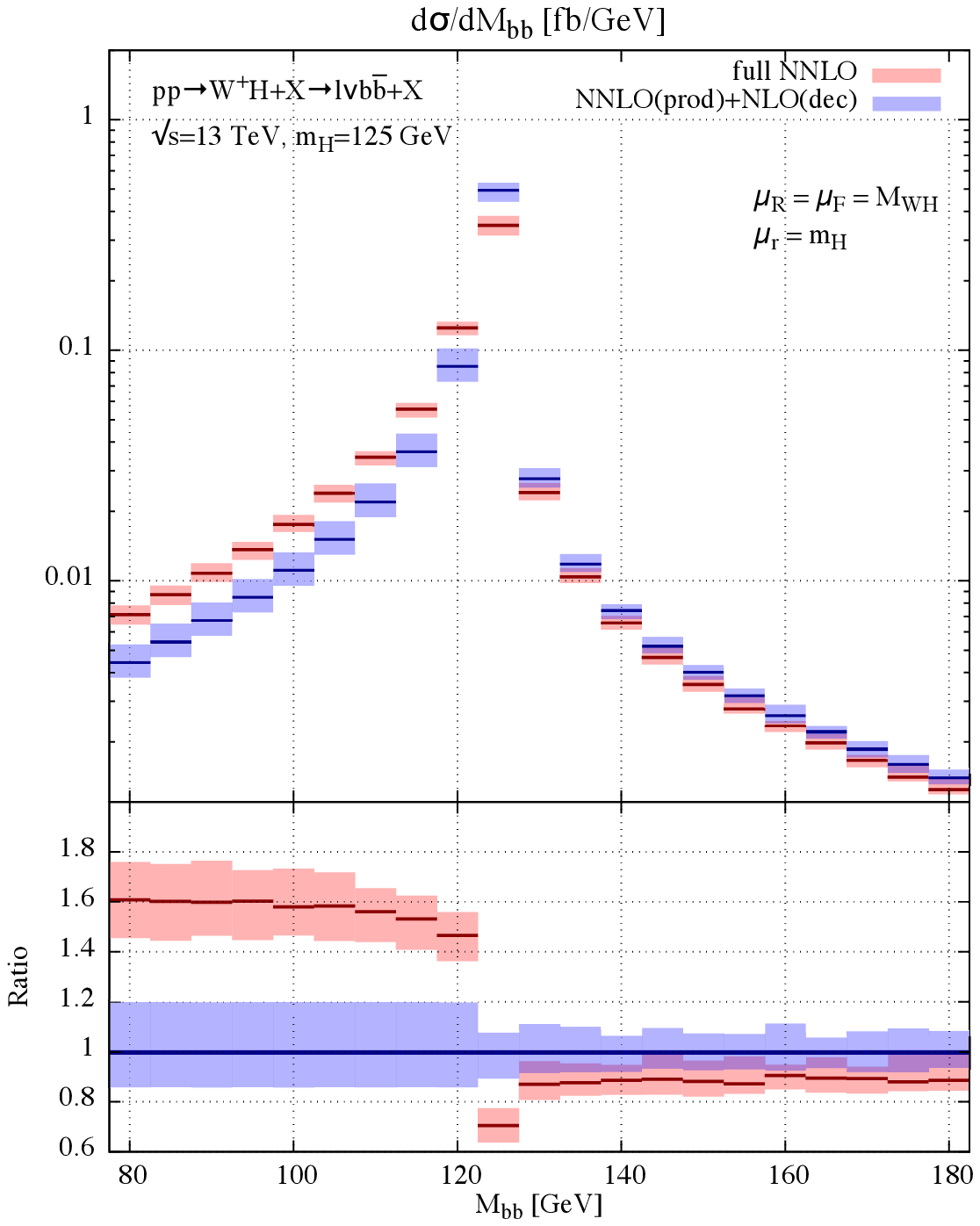}
\end{tabular}
\end{center}
\vspace*{-.3cm}
\caption{\label{fig:1}
{\em 
$pp\to W^+H+X\to l\nu_l b{\bar b}+X$ at LHC with $\sqrt{s}=13$~TeV.
Transverse-momentum distribution (left panel) and invariant mass distribution (right panel) 
of the leading $b$-jet pair computed at full NNLO (red) and partial NNLO (blue). 
The lower panels show the ratios of the results. The applied cuts are described in the text. 
}}
\end{figure}

We next consider differential distributions. In Fig.~\ref{fig:1} (left) we present the transverse-momentum distribution  $p_T^{bb}$ 
of the leading $b$-jet pair  (i.e.\ the two 
$b$-jets with largest $p_T$). 
In the lower panel  we show the ratio of the two theoretical predictions defined above. 

We observe that the additional $\alpha_S^2$ corrections included in the {\itshape full} NNLO prediction have an important effect
also on the {\itshape shape} of the $p_T^{bb}$ distribution.
In particular the cross section 
is increased by 
around $2-5\%$ for $p_T^{bb}\ltap 140$~GeV  and it is decreased by around $6-8\%$ for $p_T^{bb}\gtap 140$~GeV. 
The corresponding $K$-factor, defined as the ratio between
the {\itshape full} NNLO prediction in Eq.\,(\ref{expand2})  and the {\itshape partial} NNLO prediction in Eq.\,(\ref{expand3}),
is thus remarkably not constant (see the lower panel of Fig.~\ref{fig:1} (left)). 
The qualitative behaviour of these effects is not unexpected. The additional QCD radiation in
the Higgs boson decay, which is included in the full NNLO calculation, has the effect of decreasing the transverse-momentum of the leading $b$-jet pair, making the
$p_T^{bb}$ distribution softer.

In Fig.~\ref{fig:1} (right) we present the invariant mass distribution of the leading $b$-jet pair, $M_{bb}$.
We consider again the  comparison between the {\itshape full} NNLO QCD prediction in Eq.\,(\ref{expand2})  and the {\itshape partial} NNLO
prediction in Eq.\,(\ref{expand3}) and we show the ratio of the two predictions in the lower panel. For this observable the effect of the 
NNLO corrections to the  decay rate are even more substantial. While the position of the peak is rather stable  around the value of the Higgs boson
mass $M_{bb}\simeq m_H$, 
the spectrum receives large positive corrections (up to $+60\%$)
for $M_{bb}< m_H$ and sizeable negative corrections (from $-30\%$ to $-10\%$) for $M_{bb}\gtap m_H$.  
The large impact of these corrections can be understood by noting
that the leading order (LO) computation would produce an invariant mass distribution which exactly fulfills the constraint $M_{bb}=m_H$.
Higher-order corrections to the decay decrease the invariant mass of the leading $b$-jet pair. 
In the $M_{bb}< m_H$ region  the {\itshape partial} NNLO prediction (which contains just the NLO correction to the decay rate) is effectively a first-order 
calculation and the next-order term is contained only in the {\itshape full } NNLO correction.
Conversely, higher-order corrections to the production cross section typically increase the invariant mass of the leading $b$-jet pair and
the region $M_{bb}> m_H$ receives contributions only from partons emitted from the initial state.
In this case the effect of the additional $\alpha_S^2$ corrections contained in the {\itshape full } NNLO calculation 
has a sizeable but moderate impact with respect to the {\itshape partial} NNLO calculation.

As for the perturbative scale variation we have found that 
the scale dependence is dominated by the effect of the 
renormalization scale of the decay process $\mu_r$ and is particularly 
small: at the $1\%$ level for the fiducial cross section. The scale variation
of the ``full'' NNLO result is around $\pm 5\%$ in the case of $p_T^{bb}$ distribtution 
and around $\pm 10\%$ in the case of $M_{bb}$ distribution.
The scale dependence of the ``partial'' NNLO result is quantitatively similar being significantly larger (around $\pm 17\%$) only in the 
region $M_{bb}< m_H$, where the ``partial'' NNLO result is a first-order calculation.

We observe that the uncertainty bands for the ``partial'' and ``full'' NNLO results 
fail to overlap for the fiducial cross section and in various regions of differential distributions. 

\begin{figure}[th]
\begin{center}
\begin{tabular}{cc}
\includegraphics[width=.45\textwidth]{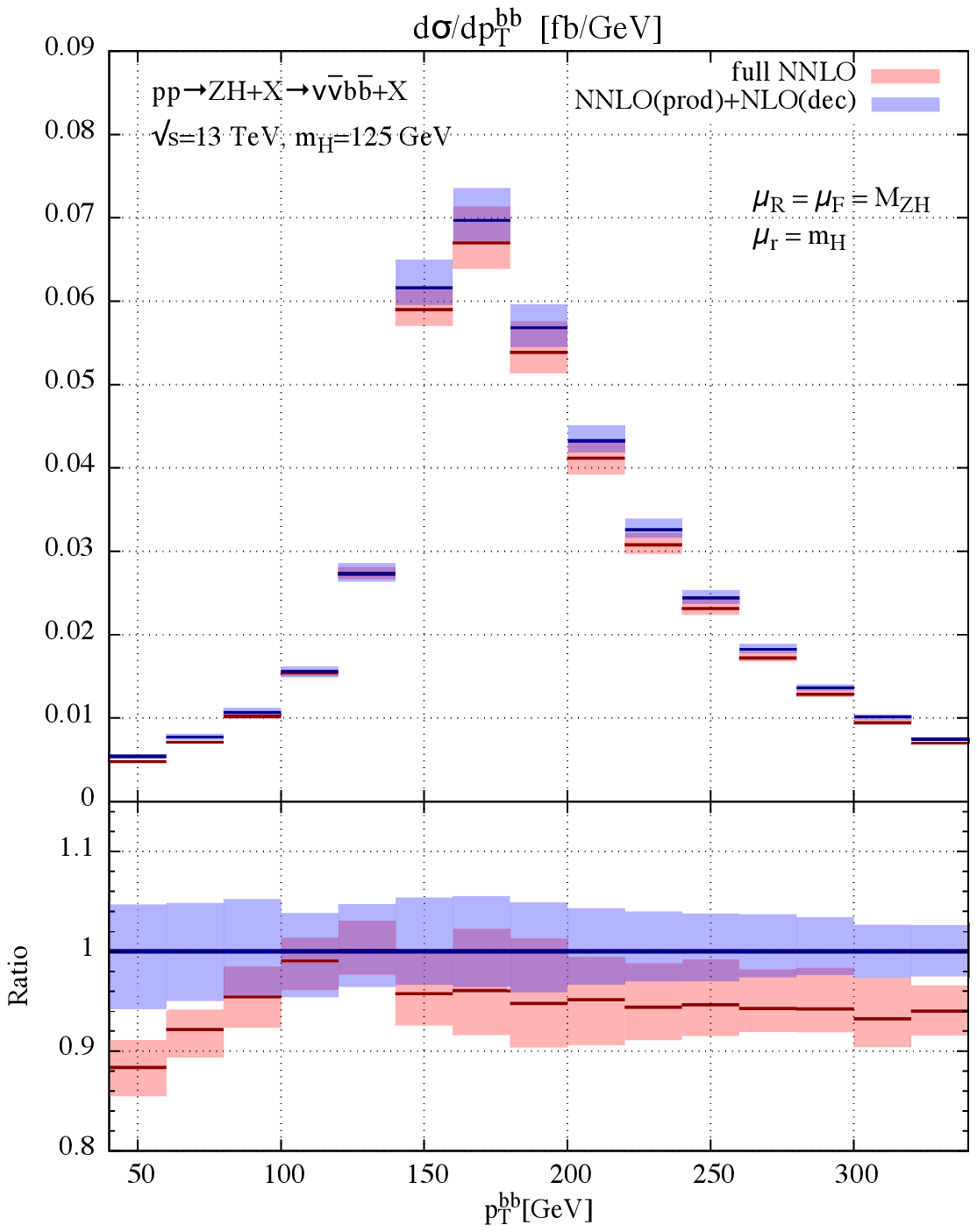}
\hspace*{.75cm}
\includegraphics[width=.45\textwidth]{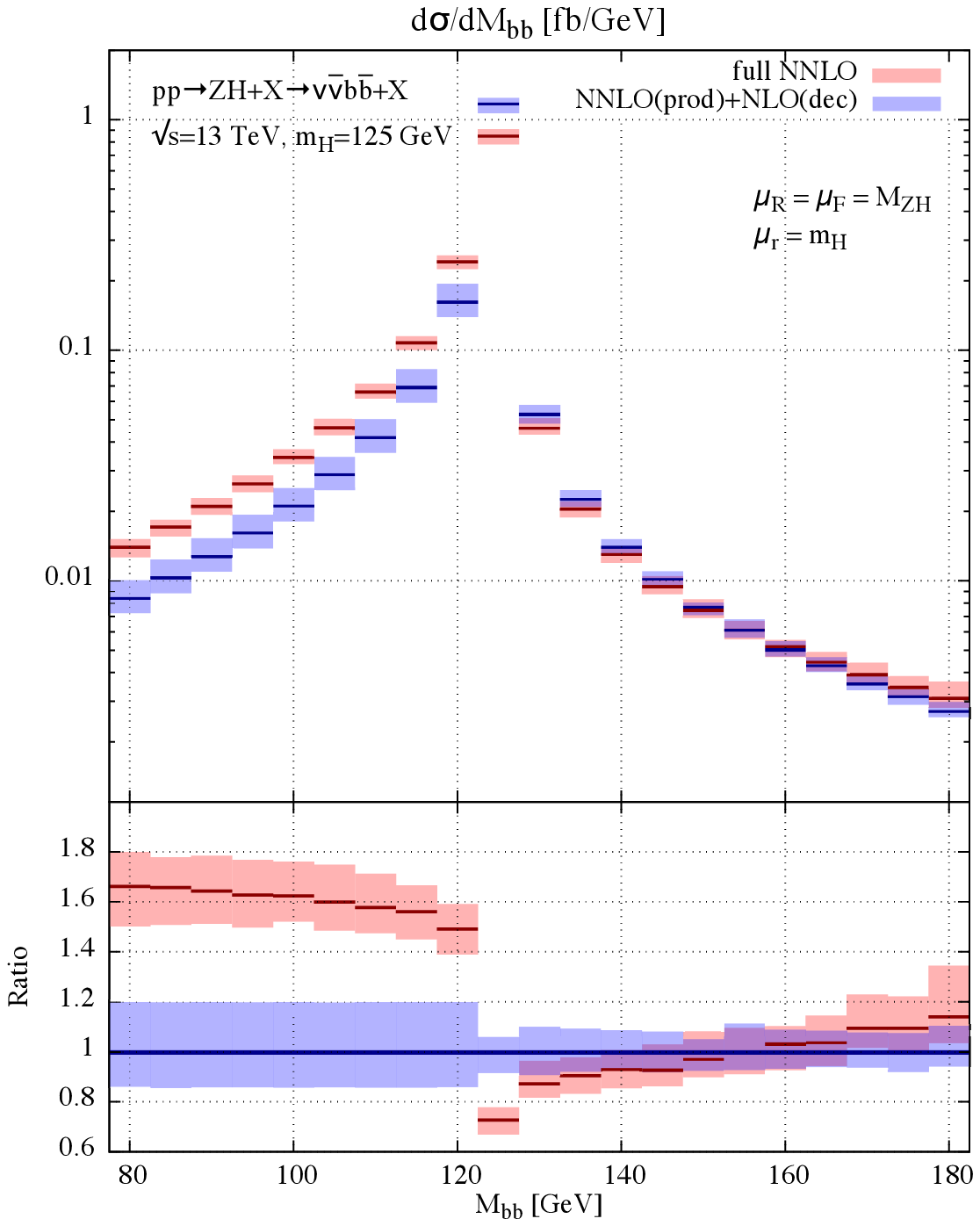}
\end{tabular}
\end{center}
\vspace*{-.3cm}
\caption{\label{fig:2}
{\em 
$pp\to ZH+X\to \nu\nu b{\bar b}+X$ at LHC with $\sqrt{s}=13$~TeV.
Transverse-momentum distribution (left panel) and invariant mass distribution (right panel) 
of the leading $b$-jet pair computed at full NNLO (red) and partial NNLO (blue). 
The lower panels show the ratios of the results. The applied cuts are described in the text. 
}}
\end{figure}

We next turn to the case of \ZH\ production and decay at the LHC at $\sqrt{s}=13$~TeV. 
We consider the invisible $Z$ decay into neutrinos ($Z\to \nu\bar{\nu}$) and we  require to have at least two $b$-jets each with $p_T^b > 25$~GeV and $|\eta_b|<2.5$ and a missing transverse energy $E_T^{miss}>150$~GeV.
The corresponding cross sections in the fiducial region are reported in the second row of Table~\ref{table1}. 
We observe that  the inclusion of the {\itshape full} NNLO corrections reduces the cross section by around $5\%$ with respect to 
the {\itshape partial} NNLO result.

In Fig.~\ref{fig:2} (left) we present the transverse-momentum distribution of the leading $b$-jet pair,  $p_T^{bb}$.
As in the previous case we compare the  {\itshape full} NNLO QCD prediction (Eq.\,(\ref{expand2})) 
with the  {\itshape partial} NNLO prediction (Eq.\,(\ref{expand3}))
and in the lower panel  we show the ratio of the two predictions. 

In this case the inclusion of 
the NNLO corrections to the decay rate decreases the cross section up to about 10\% below the peak and 
around 5\% above the peak. The corresponding $K$-factor is shown in the lower panel of  Fig.~\ref{fig:2} 
(left). 

Finally in Fig.~\ref{fig:2} (right) we consider, for the \ZH\ case,  the invariant mass distribution of the leading $b$-jet pair, $M_{bb}$.
The effect of the {\itshape full} NNLO corrections is similar to the \WpH\ case.
The spectrum receives large positive corrections (up to $+70\%$)
for $M_{bb}< m_H$ and sizeable negative corrections (from $-30\%$ to $-10\%$) for $125 \ltap M_{bb}\ltap 150$~GeV.

We finally observe that when  a kinematical boundary is present at a given order in perturbation theory, higher order
corrections are affected by instabilities of Sudakov type~\cite{Catani:1997xc},
which spoil the reliability of the fixed-order expansion around the boundary. This is the case for both the 
$p_T^{bb}$ distribution ($p_T^{V}> 150$~GeV LO kinematical boundary)
and the $M_{bb}$ distribution (LO condition $M_{bb}=m_H$). 
While a proper treatment of this misbehaviour requires an all order resummation of perturbatively enhanced terms, 
the effect of these instabilities can be mitigated by increasing the bin size of the distribution around the critical 
point. 

In the case of $ZH$ production, due to the substantial effect of the gluon-gluon initiated 
subprocess involving a heavy-quark loop, scale uncertainty is dominated by the effect of 
the renormalization scale $\mu_R$ and the ensuing scale variation band
turns out to be larger (at the $4\%$ level).  

The scale variation
of the ``full'' NNLO result is around $\pm 3-5\%$ in the case of $p_T^{bb}$ distribtution 
and around $\pm 10\%$ in the case of $M_{bb}$ distribution.
As in the \WpH\ case, the scale dependence of the ``partial'' NNLO result is significantly larger 
(around $\pm 17\%$) only in the  $M_{bb}< m_H$ region
and the uncertainty bands for the ``partial'' and ``full'' NNLO results 
fail to overlap in various regions of differential distributions. 

As already pointed out in Ref.\,\cite{Ferrera:2014lca},  we are interested 
in a specific ``boosted'' kinematical regime where the size of the NNLO
 corrections tends to be underestimated by  the customary NLO scale uncertainty 
band. Therefore the NLO scale variation cannot be regarded 
as a reliable approximation of the ``true'' perturbative uncertainty
and it casts some doubts also on the reliability of the NNLO scale variation band.

A hint on the reliability of the customary scale uncertainty at NNLO can
be obtained considering missing higher-order contributions that can be calculated
through a suitable combination of individual parts of our computation.
We have therefore calculated the $\mathcal{O}(\alpha_S^3)$ contributions
proportional to {\itshape (i)} $d\sigma^{(1)}_{h_1h_2 \rightarrow \VH} \times
d\Gamma^{(2)}_{H \rightarrow b\bar{b}}$ and {\itshape (ii)}
$d\sigma^{(2)}_{h_1h_2 \rightarrow \VH} \times
d\Gamma^{(1)}_{H \rightarrow b\bar{b}}$
in  Eq.\,\ref{expand}. We have found that the numerical impact 
to the fiducial cross-section of the N$^3$LO terms 
{\itshape (i)} and {\itshape (ii)} above
 is respectively around $-0.4\%$ and $+0.4\%$ ($-0.3\%$ and $+1.5\%$) for $WH$ ($ZH$) 
production. The fact that these effects are covered by the scale variation in 
Tab.~\ref{table1} suggests that the NNLO scale dependence (contrary to the NLO case)
could be considered as a trustable estimate of the ``true'' perturbative uncertainty
of the calculation. However a more conservative estimate of the uncertainty
can be obtained by comparing the NNLO result to what is obtained at the previous order.

We briefly comment on expected PDF uncertainties and on electroweak effects. 
PDF uncertainty has been calculated, within a similar setup, in Ref.\,\cite{deFlorian:2016spz} 
and has been shown to be at the $\pm 1.5\%$ level for fiducial cross sections.
The NLO EW effects have been calculated for $pp\to VH+X\to l_1l_2H+X$ only (i.e. without the inclusion
of EW effects for $H \rightarrow b\bar{b}$ decay)~\cite{Denner:2011id,Denner:2014cla} and it has been shown
to be significant ($\sim -10\%$)\,\cite{deFlorian:2016spz}.

In conclusion, we have presented a fully differential QCD computation for the 
associated production of a vector boson
and a Standard Model Higgs boson  in hadron collisions including the QCD 
radiative corrections up to next-to-next-to leading order
(NNLO) both for the \VH\ production cross section and for the differential 
Higgs boson decay width into bottom quarks. 
Our calculation also includes the leptonic decay of the vector boson with 
finite-width effects and spin correlations
and it is implemented in the  parton level Monte Carlo numerical 
code {\ttfamily HVNNLO}.


We have studied the impact of the {\itshape full} NNLO QCD corrections to the \VH\ production and decay 
at the LHC by focusing on the most relevant distributions, namely the  transverse momentum and invariant mass of the Higgs boson candidate. 
We have studied the renormalization and factorization scale dependence of the results in order to estimate the
perturbative uncertainty of our predictions.
We have found that the additional second-order corrections included in the present calculation have a substantial effect both for 
\WH\ and \ZH\ production. Therefore the inclusion of these effects turns out to be essential in order to obtain a precise 
theoretical prediction for associated \VH\ production and decay at the LHC.


\paragraph{Acknowledgements.} 

We gratefully acknowledge Massimiliano Grazzini for discussions and comments on the manuscript,
Raoul R\"ontsch for comparison with the results of Ref.\,\cite{Caola:2017xuq},
and the computing resources provided by the SCOPE Facility at the University of Napoli
and by the Theory Farm at the University of Milano and INFN Sezione di Milano.
This research was supported in part by Fondazione Cariplo under the grant number 2015-0761, 
by the Italian Ministry of Education and Research (MIUR) under project number 2015P5SBHT and 
by grant K 125105 of the National Research, Development and Innovation Fund in Hungary.




\end{document}